# Error reduction in biosensors using secondary labeling

Tuhin Chakrabortty[*], Manoj M Varma[*#]

[*] Center for Nano Science and Engineering, Indian Institute of Science, Bangalore
[#] Robert Bosch Centre for Cyber-Physical Systems, Indian Institute of Science, Bangalore

**Abstract:**
In this article, we derive the limit of detection for a two-step molecular recognition process and show that in-spite of all the recognition reactions being in equilibrium the overall error rates can be reduced exactly as much as possible in non-equilibrium methods such as kinetic proof-reading.

**Introduction:**

Biological systems can operate with extraordinary precision. Single cells can detect a change in concentration as low as 3.2nM [1]. DNA polymerase in some bacteria operates with a mutation rate of 1 error per 100 million to 1 billion nucleotides [2]. Surprisingly, these systems achieve this in an extremely noisy environment, where the concentration of the molecules of interest can be several orders of magnitude lower than the other molecules present in the environment. Mathematical analysis of these systems suggest that it is theoretically impossible to achieve such precision with equilibrium biochemical reactions [3]. To address this discrepancy between theory and experimental evidence, a biochemical proofreading scheme was proposed by J.J. Hopfield [4]. In his seminal work, Hopfield argued that addition of a non-equilibrium step in the biochemical process can significantly improve the specificity of the system and therefore, improve the precision.

Similar to biological systems, artificially engineered biomarker detection systems (clinical biosensors) also struggle with the presence of spurious molecules in bodily fluids. Typically, these clinical biosensors have a sensing surface coated with receptor molecules which are obtained in the form of antibodies produced in vivo against the biomarker of interest. Therefore, the binding affinity of the receptors towards the biomarker is higher than other biomolecules in the serum. However, because of the in-vivo selection process, these receptors are also often cross-reactive. Even though these antibodies are screened only against the biomarker of interest, in many cases, they have binding affinity towards several unrelated antigens present in the serum [5]. Therefore, in a biomarker detection process one also needs to consider the noise due to the binding of spurious molecules to the receptors.

In a previous work [6], we have calculated the limit of detection of a clinical biosensor in an experimentally relevant situation. Our analysis showed that if the concentration of the receptors is not known to infinite precision, then the choice of the read-out technique has a large effect on the performance of the biosensor, particularly the fidelity of measurement. With uncertainty in the receptor concentration, a read-out technique with minimum signal background from the unbound receptors performs better. This result explains the lower *LoD* of fluorescence-based techniques such as ELISA compare to label-free techniques. With an optimal read-out technique which minimizes the background from unbound receptors, the performance of a biosensor is then limited by the fluctuation in the concentration of the spurious molecules present in the serum. The limit of detection ($LoD$) of such a system can be written as

$$LoD = \sqrt{2}q\delta_n \tag{1}$$

Experimental evidence suggests that this limit of detection can be significantly improved by adding a second step where a label molecule is introduced into the system. Similar to the receptors, these label molecules have a higher binding affinity towards the biomarker of interest and therefore selectively binds to them. Interestingly, addition of the second molecule in these system mimics the effect of proofreading in cellular sensing systems, even though, unlike Hopfield's proofreading schemes, all the reactions in case of the two-step diagnostic processes are in equilibrium. In this article, we derive a general expression for the limit of detection of a two-step diagnosis process. In the light of the results in our previous work, we will then limit ourselves to the analysis of fluorescence-based two-step detection techniques to study the effects of addition of a label molecule in the sensing system. We will conclude our analysis by comparing a two-step diagnosis process with traditional non-equilibrium proofreading schemes.

**Mathematical model**

Let us consider a two-step biomolecular detection system using a fluorescent label even though the approach here is applicable to any other labels such as enzymes in the case ELISA. The first step is to pour the serum over a surface where receptors are immobilized. Ideally, this should cause biomarkers to specifically attach to the surface. However, non-specific antigens present in the serum also get attached to the surface due to the low specificity of the receptors. Therefore, in the second step, a fluorescently labeled antibody is applied over the surface. Similar to [6], to detect the presence of specific biomarkers in the serum, a baseline measurement is done with healthy serum and is compared with the serum to be tested. If $S$ and $S_0$ are the output signals from experiments in presence and absence of specific ligands respectively, then the generic expressions for $S$ and $S_0$ can be written as

$$\begin{aligned} S &= \rho_b^{(2)} b^{(2)} + \rho_b^{(1)}\big(b^{(1)} - b^{(2)}\big) + \rho_u\big(C_r - b^{(1)}\big) + \xi \\ S_0 &= \rho_b^{(2)} b_0^{(2)} + \rho_b^{(1)}\big(b_0^{(1)} - b_0^{(2)}\big) + \rho_u\big(C_r - b_0^{(1)}\big) + \xi \end{aligned} \tag{2}$$

Where $C_r$ is the concentration of the receptor molecules, $b^{(i)}$ and $b_0^{(i)}$ are the fractions of bound receptors in present and absent of specific ligands and $\rho_b^{(i)}$ and $\rho_u$ are the signal strength due to bound and unbound receptors respectively. The superscripts 1 and 2 represent the 1st and 2nd step respectively. For example, $b^{(2)}$ is fraction of receptors which are bound with the labels and $b^{(1)}$ is the fraction of receptors which are only bound with the ligands but not with the label. Similarly, $\rho_b^{(1)}$ and $\rho_b^{(2)}$ represent the signal due to the receptors which are bound only by the ligand molecules and receptors which are bound with both ligand molecules and labels respectively. The parameter $\xi$ is the measurement noise, which is defined as $\xi = \mathcal{N}(\mu_\xi, \sigma_\xi^2)$.

$\mu_\xi$ is the average fluorescence due to the binding of the label molecules to the sensor surface and $\sigma_\xi^2$ is the variance in the measured parameter due to the noise in the measurement system.

The parameters $b^{(2)}$ and $b_0^{(2)}$ can be defined as

$$b_0^{(2)} = \left(\frac{C_r C_n}{C_r + K'_{Dn}}\right)\left(\frac{C_m}{C_m + K_{Dn}}\right)$$

$$b^{(2)} = \left(\frac{C_r C_n}{C_r + K'_{Dn}}\right)\left(\frac{C_m}{C_m + K_{Dn}}\right) + \left(\frac{C_r C_s}{C_r + K'_{Ds}}\right)\left(\frac{C_m}{C_m + K_{Ds}}\right) \quad (3)$$

where $K_D$ is the dissociation constant for the ligand-fluorescent label interaction, $K'_D$ is the dissociation constant for the ligand-receptor interaction and the subscripts $r, s, n$ and $m$ represent receptors, specific ligands, non-specific ligands and labels respectively. Typical biosensors have receptor concentration much larger than the specific and non-specific ligand concentrations to avoid undetectability in presence of measurement noise [6]. Therefore, the expression for $b^{(2)}$ and $b_0^{(2)}$ in equation (3) can be rewritten as

$$b_0^{(2)} = C_n \left(\frac{C_m}{C_m + K_{Dn}}\right)$$

$$b^{(2)} = C_n \left(\frac{C_m}{C_m + K_{Dn}}\right) + C_s \left(\frac{C_m}{C_m + K_{Ds}}\right) \quad (4)$$

Similar to [6], to consider the effects of the fluctuations in the concentrations of receptor and non-specific ligand molecules, we have assumed $C_n$ and $C_r$ as Gaussian distributions defined as

$$C_r = \overline{C_r}(1 + \mathcal{N}(0, \delta_r^2))$$
$$C_n = \overline{C_n}(1 + \mathcal{N}(0, \delta_n^2))$$

In a real-life scenario, the concentration of the fluorescent labels depends on the probability that a label molecule is tagged with the fluorescent molecules. However, it is always possible to eliminate the noise generated by the stochasticity of tagging by selecting only the tagged molecules. Therefore, without loss of generality, we can safely assume the concentration of label molecules ($C_m$) to be constant. Therefore, following the derivation in [6], the *LoD* of a two-step biomarker sensor can be written as

$$LoD^{(2)} = \sqrt{2}q \left(\frac{\gamma_n}{\gamma_s}\right) \sqrt{\left(\left(\frac{\rho_u \overline{C_r}}{\gamma_n \overline{C_n}}\right)^2 \delta_r^2 + \delta_n^2\right)} \quad (5)$$

Where $LoD = \left(\frac{C_s}{C_n}\right)_{min}$. $\delta_i = \sigma_i/\overline{C}_\iota$, $\gamma_i = \left(\rho_b^{(2)} - \rho_b^{(1)}\right)\left(\frac{C_m}{C_m+K_{Di}}\right) + \left(\rho_b^{(1)} - \rho_u\right)$ and $q = Q^{-1}(PE^{(\infty)}/2)$, where $Q$ is the $Q$-function and $PE^{(\infty)}$ is the probability of error with infinite measurements.

**Results and discussion:**

We can retrieve the limit of detection of a one-step detection system [6] from equation (5) for $C_m = 0$. Similar to the one-step detection system, the limit of detection in case of a two-step system $(LoD^{(2)})$ also explicitly depends on the parameters $q, \delta_r, \delta_n$ and $\rho_u$. As a result, the effects of these parameters on the limit of detection remain unchanged after the addition of the second step. In contrast, the parameter $\left(\rho_b^{(1)}\right)$ has exactly the opposite effect than it had in the one-step system. To achieve a lower limit of detection in case of a one-step system, $\left(\rho_b^{(1)}\right)$ needed to be high. Instead, in case of a two-step system, optimal limit of detection can be achieved by minimizing $\left(\rho_b^{(1)}\right)$ while maximizing $\left(\rho_b^{(2)}\right)$. [Fig 1a] Therefore, for an optimal two-step biomarker detection system, $\rho_u = \rho_b^{(1)} = 0$. For such a system, the limit of detection becomes

$$LoD_{opt}^{(2)} = \sqrt{2}q\chi\delta_n \tag{6}$$

Where $\chi = \frac{C_m+K_{Dn}}{C_m+K_{Ds}}$ and $\delta_n = \sigma_n/\overline{C}_n$

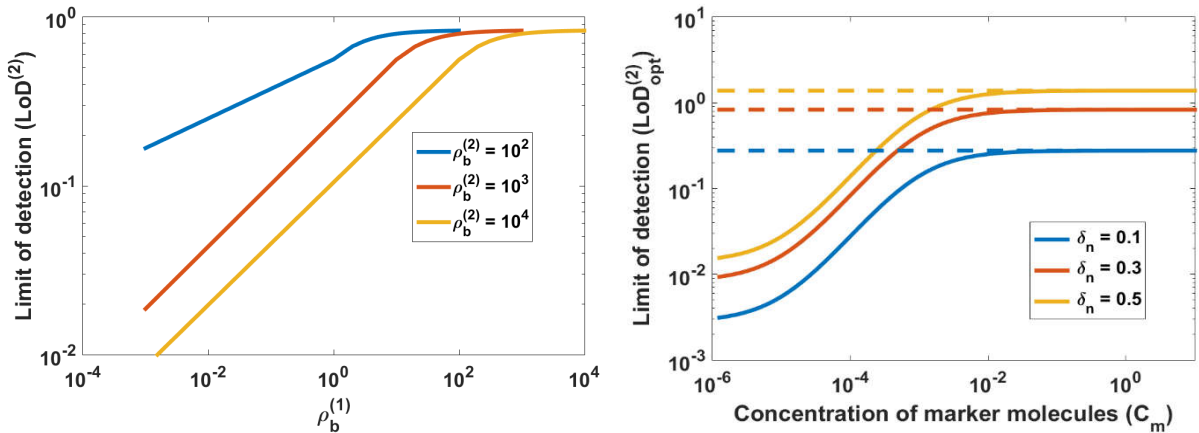

[Figure 1: a. Effect of $\rho_b^{(1)}$ on the limit of detection. For the case of a two-step diagnosis process, decreasing $\rho_b^{(1)}$ reduces the *LoD* and hence makes the system more efficient. [Parameters used: $\rho_u = 0.001, C_m = 0.5 \times 10^{-7}, K_{Ds} = 10^{-5}, K_{Dn} = 10^{-3}, C_r = 10^{-2}, C_n = 10^{-4}, PE = 0.05, \delta_r = \delta_n = 0.3$] b. Effect of concentration of label molecules on the optimal limit of detection $(LoD_{opt}^{(2)})$ for a two-step system for different values of $\delta_n$. The dotted lines represent the limit detection for a one-step system for the same values of $\delta_n$. Introducing a second step improves the limit of detection of a system for very small concentration of label molecules. Interestingly, increasing the concentration of the label molecules increases the limit of detection and eventually converges with the limit of detection of a one-step system for very large concentration of label molecules. Parameters Used: $PE = 0.05, K_{Ds} = 10^{-5}, K_{Dn} = 10^{-3}$ ]

Figure (1b) shows the effect of $C_m$ on the limit of detection. One interesting aspect of equation (6) is that $LoD_{opt}^{(2)}$ converges to the $(LoD)_{opt}$ calculated in equation (10) of [6] for $C_m \gg K_{Dn}$. Therefore, adding a large concentration of label molecules doesn't increase specificity of the optimal detection system. One could intuitively interpret this in the following manner. In case of $C_m \gg K_{Dn}$, because of the abundance of the fluorescent labels, there will be no competition between the specific and non-specific ligands for binding to the labels. Therefore, the labels will bind to all the binding sites with equal probability and the assay will get saturated. On the other hand, addition of label molecules in very low concentration can improve the $LoD$ significantly. Comparing $LoD_{opt}^{(2)}$ with $LoD_{opt}$ derived in [6] for $C_m \ll K_{Dn}$, we get

$$LoD_{opt}^{(2)} = K_D LoD_{opt}^{(1)} \qquad (7)$$

We can compare this result with the kinetic proofreading scheme described by Hopfield [4]. Hopfield demonstrated that the probability of formation of an error product of an equilibrium Michaelis-Menten system can be improved by addition of a non-equilibrium step. His results suggest that the maximum possible improvement in specificity in a such a system will be of the order of $K_D$, where $K_D$ is the ratio of the dissociation constants of the correct and error reactants. Although, results in equation (7) is similar to Hopfield's results, unlike Hopfield's model, all the reaction kinetics in this case are in equilibrium. One can consider the addition of the label molecule into the system as the out-of-equilibrium step in the process as it breaks the law of conservation of mass. In fact, addition of the label molecule provides additional information and thereby decreases the entropy of the system. Therefore, the two-step diagnosis system doesn't need a non-equilibrium step to achieve a higher specificity.

**Conclusion:**

In [6], we had calculated the limit of detection of a one-step biomolecular detection system. We showed that under optimal experimental conditions, the $LoD$ is limited by the fluctuation in the concentration of the non-specific ligand molecules present in the serum. In this article we extended the analysis to two-step detection systems where a label molecule is added to the system as a second step. Our results show that addition of label molecule in small concentration can reduce the $LoD$ significantly as observed in commercially available systems such as ELISA. Surprisingly, addition of large concentration of label molecules results in no measurable change in the $LoD$ of the system.

Our analysis provides a mathematical framework for designing two-step biomarker detection systems. We demonstrated that for optimal two-step systems, the contrast between the signals due to the unbound receptors and receptors bound only in the first step of the process needs to be minimized. Therefore, detection systems that perform poorly as one-step systems, can be excellent candidates for a two-step detection system.